\documentclass[aps,prl,showpacs,twocolumn]{revtex4}
\usepackage{dcolumn}

\begin{document}
\pagestyle{empty}

\textbf{Bakhtiari, Leskinen and T\"orm\"a Reply:} The Comment \cite{Molina2009a} displays calculations showing the expected result that, for the specific case considered, the oscillations predicted by DMRG studies are smaller than those predicted by mean field calculations. We emphasize that the aim of our Letter \cite{Bakhtiari2008a} is not to provide accurate predictions for the density and order parameter profiles for this particular system. 

The Comment points out the need to clarify that the aim and the main result of our Letter actually was to show how rf spectroscopy reveals information about an oscillating order parameter. In particular, we argue that \textit{if} the order parameter oscillates, the Andreev states that it may host at its nodes manifest as contributions at negative detunings in rf spectroscopy, due to momentum conservation. This argument is general and does not depend on whether the order parameter oscillations are actually strong in the specific 1D example case that was considered in order to illustrate the argument. Obtaining not only the ground state but also the excitation spectrum, necessary for calculating the rf spectrum, is highly demanding within DMRG. Therefore a mean field approach is a reasonable first choice, especially when it is used for demonstrating a principle, not for quantitative predictions on a particular system. Although the difference between the DMRG and mean field orde
 r parameter profiles in this particular example case may be called qualitative, this has no effect on the qualitative arguments that we present about rf spectroscopy. 

In other words, we could have also formulated the problem in the following way: \textit{Assume} that some physical system has an excitation spectrum associated with a strongly oscillating order parameter -- how is that visible in rf spectroscopy? This is the question that our Letter answers. Note that the Letter already shows, via the finite temperature study, that if the oscillations decrease, also the effect on the rf spectrum becomes smaller. It is a subject of further study to model experimentally realistic, perhaps higher dimensional and larger systems and estimate whether the predicted signal in the rf spectum is within the experimental resolution. The main point of the present Letter was to suggest using a mean-field analysis how and why the rf-spectrum displays signatures of an oscillating order parameter, in case considerable oscillations occur in the system.   \\
\\
M. Reza Bakhtiari, M.J. Leskinen and P. T\"orm\"a,
Department of Applied Physics, P.O. Box 5100, 02015 Helsinki University of Technology, Finland


\begin{thebibliography}{77}

\bibitem{Molina2009a} R.A. Molina, J. Dukelsky and P. Schmitteckert, \prl {\bf 102}, 168901 (2009).

\bibitem{Bakhtiari2008a} M.Reza Bakhtiari, M.J. Leskinen, P. T\"orm\"a, \prl {\bf 101}, 120404 (2008).

\end{thebibliography}
\end{document}